\documentstyle[prb,preprint,aps]{revtex}
\begin{document}
\draft
\preprint{\footnotesize{Applied Physics Report 94-35}}
\title{Influence of Long-Range Coulomb Interactions
on the Metal-Insulator Transition in One-Dimensional
Strongly Correlated Electron Systems}
\author{I.~V.~Krive$^{(1,2)}$, A.~A.~Nersesyan$^{(1,3)}$,
M.~Jonson$^{(1)}$, and R.~I.~Shekhter$^{(1)}$}
\address{
 $^{(1)}$Department of Applied Physics, Chalmers University of
Technology and G\"{o}teborg University, S-412 96 G\"{o}teborg,
Sweden }
\address{
$^{(2)}$B. I. Verkin Institute for Low Temperature Physics and Engineering,
47 Lenin Avenue, 310164 Kharkov, Ukraine
}
\address{
$^{(3)}$Institute of Physics, Georgian Academy of Sciences,
Tamarashvili 6, 380077, Tbilisi, Georgia}
\maketitle

\begin{abstract}
The influence of long-range Coulomb interactions on the
properties of one-dimensional (1D) strongly correlated
electron systems in vicinity of the metal-insulator phase
transition is considered. It is shown that unscreened repulsive
Coulomb forces lead to the formation of a 1D Wigner crystal
in the metallic phase and to the transformation of the square-root
singularity of
the compressibility (characterizing the commensurate-incommensurate
transition) to a logarithmic singularity. The properties
of the insulating (Mott) phase depend on the character of the
short-wavelength screening of the Coulomb forces. For a sufficiently
short screening length the characteristics of the charge excitations
in the insulating phase are totally determined by the Coulomb interaction and
these quasipartic les can be described as quasiclassical Coulomb solitons.
\end{abstract}
\newpage

\sloppy

The metal-insulator transition induced by strong correlations in an
electron system
is a problem of permanent theoretical interest.
In recent years, this problem became particularly important in connection
with the discovery of high-temperature superconductivity and subsequent
attempts to develop a consistent theory of this phenomenon [1].

Much progress has been achieved in one-dimensional models where exact results
or well developed non-perturbative methods are available,
making it possible to obtain an
analytical description of the dynamics of the Mott-Hubbard transition in a wide
range of bare parameters of the system. For example, using
the Bethe-ansatz solution of the 1D Hubbard model, an analytical expression
has been recently obtained for the charge
stiffness of a finite-size system close to the transition as a function of the
on-site repulsion $U$ [2].

It is well known that, at half filling (one electron per site), the 1D
repulsive Hubbard model describes a Mott insulator. In this case,
the charge excitations have a gap in their spectrum and at weak
Hubbard interaction, $U \ll t$ ($t$ being the bandwidth), can be  regarded as
topological solitons (kinks) of the quantum sine-Gordon model with coupling
constant $\beta^2 = 8 \pi$ (see {\it e.g.} [3]). At finite deviations from half
filling the system is in a metallic phase, but the charge carriers
(``holons") keep the memory of the Mott phase; they can still be considered
as solitons with a characteristic size proportional to the correlation length
$\xi_0$ in the insulating phase. At low densities, $n \xi_0 \ll 1$, the holons
interact weakly and can be described as free massive spinless fermions [2,3].

In this paper we study the effects of a long-range Coulomb interaction on the
metal-insulator transition in strongly correlated one-dimensional
electron systems. We shall
assume that the Mott phase for the charge excitations is described in terms of
the sine-Gordon model, while the transition to the metallic phase is related to
creation of solitons at the chemical potential exceeding the charge gap.
Near the phase transition,  interactions between the solitons can be
neglected,  so that in the absence of Coulomb correlations the
properties of the system near the transition are completely determined by Fermi
statistics of quantum solitons. In particular, the compressibility of the
system has a trivial (in fermionic language) square-root singularity typical
for the commensurate-incommensurate transition [4,5,6]. Notice that, within
a classical description, a system of kinks with a finite density
forms a soliton lattice at zero temperature (see, {\it e.g.} [7]).
However, the
interaction between the solitons is short ranged and, at densities
$n \xi_0 \ll 1$, exponentially small. Quantum fluctuations, representing the
Goldstone mode of the soliton lattice, destroy the periodic structure.
Therefore, irrespective of the interaction constant, the transition to the Mott
phase always occurs from a disordered phase, the latter described in terms of
free spinless fermions. When the long-range Coulomb interaction
is taken into account, the above described
qualitative picture is modified. The unscreened repulsive Coulomb forces
between the solitons lead to the formation of a Wigner crystal [8]. Strictly
speaking, in a  1D infinite chain of charges the $1/r$-interaction does not
fall off fast enough to remove the infrared divergence in the density-density
correlation function. However, the $4k_F$ oscillating part of the correlator
decays slower than any power [9], thus making it reasonable to speak of a
nearly
ordered state. At low densities $n a_B \ll 1$ ($a_B = \hbar^2 / e^2 m_s$
and $m_s$ being the Bohr radius and soliton mass, respectively) the Wigner
lattice with perod $a \equiv n^{-1}$ is not destroyed at exponentially large
distances $L \leq a \exp (const /n a_B)$. Therefore, one can expect,
in finite-size (mesoscopic) samples, the "holons" to form a Wigner crystal, so
that the transition to the insulating phase takes place from the Wigner crystal
phase. We show that, in this case, the square-root singularity of the
compressibility is changed to a logarithmic singularity.

The holon density dependence of the behavior of the Wigner lattice is
determined by the screening of the Coulomb interaction both
at small and large distances. In quasi-one-dimensional systems, the ultraviolet
cutoff is determined, as a rule, by the characteristic width $\lambda$ of
a "one-mode" channel. Long-distance screening can occur
when the one-dimensional chain of length $L$ is situated close to a large
metallic electrode, at at distance $D \ll L$. In this case the behavior of the
compressibility $\kappa = \partial n /\partial \mu$
in the weak-screening $(n D \geq 1)$ and strong-screening
$(n D \ll 1)$ regimes turns out to be qualitatively different.
At weak screening, the dependence $\kappa = \kappa (n)$ does not differ from
that in the case of an unscreened Coulomb interaction (with the obvious change
$L \rightarrow D$): $\kappa (n) \sim 1 / \ln (nD)$. On lowering the density,
one passes to the strong screening regime
$n D \ll 1$, where the above logarithmic behavior is replaced by a power-law:
$\kappa (n) \sim (Dn)^{-2}$. In the limit of arbitrarily small densities, this
dependence would lead to the exponent 2/3 in the dependence of
the compressibility on the chemical potential near the threshold: $\kappa (\mu)
\sim (\mu - \mu_c)^{-2/3}$. However, in the vicinity of the phase transition,
quantum fluctuations melt the Wigner lattice and recover the universal
square-root singularity $\kappa (\mu) \sim (\mu - \mu_c)^{-1/2}$,
characterizing
the commensurate-incommensurate transition [4]. The presence of a large
metallic electrode allows the range of quantum fluctuations to be changed.
On the other hand, it
also leads to an additive renormalization of the critical value of $\mu$,
$\mu_c = \Delta - e^2 / 4D$, due to effects caused by image forces.

The magnitude of the gap $\Delta$ in the spectrum of charged excitations in the
insulating phase crucially depends on the character of the short-wavelength
screening of the Coulomb interaction. This can be easily understood for
weak Hubbard interaction $U \ll t$, when the charge excitations are quantum
solitons with characteristic correlation length $\xi_0 \sim \hbar c_s /\Delta$
($c_s$ is the velocity of the  charge excitations in the Mott phase).
If $(\xi_{0} / \lambda) \alpha _s \leq 1$ $(\alpha _s = e^2 / \hbar c_s)$, the
long-range Coulomb interaction weakly affects characteristics of the
insulating phase, and the charge gap is determined by the well-known formula of
Lieb and Wu [10].

When the opposite inequality is realized, $(\xi_0 / \lambda) \alpha_s \gg 1$,
the Coulomb forces significantly renormalize (increase) the Mott-Hubbard gap.
It is physically evident that in this case the charge excitations
--- solitons ---
become heavier $(\Delta_C \gg \Delta)$ due to the strong electrostatic energy,
and ``fragile" ($\xi_C \gg \xi_0$) due to the unscreened Coulomb repulsion of
charges ``inside" the soliton. It seems reasonable to call these excitations
Coulomb solitons, since their behavior in the conduction band at
$\mu > \Delta_C$
differs from that described above. Namely, at low densities, $n \xi_s \ll 1$,
the Coulomb solitons still condense into a Wigner crystal and display the
logarithmic threshold singularity of the compressibility. However, at densities
$n \xi_s \geq 1$, but still much lower than $a^{-1} _B$, the Wigner lattice is
transformed to a sine-Gordon soliton lattice. The compressibility saturates
at values $\kappa \sim (\Delta_C \xi_s)^{-1}$, characteristic
for a charge-density wave.

Let us consider first the influence of long-range Coulomb forces on the
properties of the metallic phase near the transition to the insulating state.
In what follows, we shall assume that the transition is of the Mott-Hubbard
type, although our conclusions remain valid for any one-dimensional charged
system near the commensurate-incommensurate transition.

Using the standard approach (see {\it e.g.} [11], and also [12]), one can treat
the charge excitations of the half-filled, weak-coupling ($U \ll t$) Hubbard
model as topological quantum solitons of the related sine-Gordon model
(at $\beta^2 = 8\pi$), characterized by correlation length
$\xi_0 = \hbar c_s/\Delta$ (here $2\Delta$ is the Mott-Hubbard gap, and
$c_s \simeq 2t + U/2\pi$ is the velocity of charged excitations). The
long-range repulsive forces will be taken into account by adding the screened
(at short distances $\Delta x \ll \lambda$) Coulomb interaction energy to the
total energyof
 the system. The screening length $\lambda$ is determined, for example,
by transverse dimensions of the confinement potential which forms the
one-dimensional chain.

Let $\lambda \gg \xi_0$. Then it is reasonable to assume (the exact criterium
will be given below) that the Coulomb interaction does
not affect the intrinsic characteristics of the solitons ({\it i.e.}
the Mott phase) but changes the
character of interaction between them at large distances,
$\Delta x \geq \lambda$.
Therefore, assuming the solitons to be point-like objects, we shall study
their dynamics in the conduction band at low densities,
$n \ll a^{-1}_B$. At densities $n \leq a^{-1}_B$
 the Coulomb
interaction leads
to the formation of a soliton Wigner lattice with period $a \equiv n^{-1}$
 ($a_B = (\hbar c_s / e^2) \xi_0$ is the Bohr radius
for charged particles with mass $m_s = \Delta/c^2 _s$).
This ordered state is expected to be stable in chains of mesoscopically large
length and therefore can significantly affect the properties of the metallic
phase [13]. We shall study the influence of Wigner crystallization on the
characteristics ofth e transition to the insulating (Mott) phase.

Consider a situation, typical for ``mesoscopic" experiments, when a massive
metallic electrode is placed at the distance $D$ ($\lambda \ll D \ll L$) from
the chain. The electrode leads to screening of the Coulomb interaction at large
distances. In this case, the classical (electrostatic) energy of the lattice,
with the electrode screening effect taken into account, equals [13]
\begin{eqnarray}
{\cal E}_W &=& \frac{e^2 n^2}{2} \left[ \sum_{i \neq 0} \frac{1}{|i|}
- \sum_{i=-\infty}^{\infty} \frac{1}
{\sqrt{i^2 + (2 D/a)^2}} \right]\nonumber\\
&=& e^2 n^2 \left[ \ln (n D) + C - 2 \sum_{k=1}^{\infty}
K_0 (4 \pi k n D)
\right] ,
\end{eqnarray}
where $C$ is the Euler constant, and $K_0 (x)$ is the Macdonald function.
The asymptotic behavior of the expression (1) in the limits of high and low
densities are
\begin{eqnarray}
{\cal E}_W &\simeq& e^2 n^2 \ln (n D), ~~n D \geq 1\\
{\cal E}_W &\simeq& n \left( - \frac{e^2}{4D}\right)
+ 2 \zeta(3) (e D)^2 n^4, ~~n D \ll 1 .
\end{eqnarray}
Quantum corrections to the energy of the Wigner crystal (1) caused by
zero-point fluctuations
\begin{equation}
\Delta {\cal E}_W = \frac{\hbar}{2}s \int_{0}^{\pi/a} k~dk,~~~~~
s^2 = \frac{n}{m_s} \frac{d^2 {\cal E}_W}{d n^2}
\end{equation}
are small,
$\Delta {\cal E}_W \ll {\cal E}_W$ ($s$ is the sound velocity for the Wigner
crystal).  This is true at all densities $n \ll  a^{-1} _B$ in the weak
screening regime, and within the interval $a_B D^{-2} \ll n \ll a^{-1} _B$ at
strong screening ($nD \ll 1$) as well. As could be expected, a strongly
screened
Coulomb interaction at low densities $n \leq a_B D^{-2}$ cannot stabilize the
Wigner lattice which is already  melted at zero temperature by zero-point
fluctuations. In the absence of the macroscopic electrode, the regime Eq.(2) is
realized, where $D$ is replaced by the longitudinal length of the chain $L$.

As already mentioned there is no true Wigner crystallization in an infinite
chain. Nonetheless, in mesoscopic-size  systems, Coulomb correlations support
almost perfect long-range order [9],  making it possible to consider Wigner
crystallization in the strict sense. Note that in the
absence of impurities producing pinning of the Wigner crystal, the soliton
lattice can freely slide along the one-dimensional channel. Therefore, in the
ideal system, the response of a Wigner soliton-crystal to a low-frequency
electric field is identical to that of a Fermi gas of solitons. (For instance,
the charge stiffnesses are identical [14]). On the other hand, it is clear
that the compressibilities of the two systems are different, so that, upon
crystallization of solitons, the nature of the singularity at the transition to
the insulating phase on changing the chemical potential is changed.

For the Fermi gas of solitons (with the rest energy $\Delta$ and mass $m_s =
\Delta/ c^2 _s$), the thermodynamic potential of the system $\Omega$,
calculated in the low-density limit, when the solitons can be treated as
noninteracting particles, equals
\begin{equation}
\Omega_F (n) = (\Delta - \mu)~n + \frac{\pi^2 c^2 _s}{6 m_s}~ n^3 .
\end{equation}
Equation (5) leads immediately to the square-root singularity in the
equilibrium
density of holons near the phase transition to the insulating phase [5,6]:
\begin{eqnarray}
n(\mu) &=& 0, ~~~~~~~~~~~~~~~~~~~~~~~~\mu < \Delta \\
n(\mu) &=& \frac{2}{\pi} \frac{m_s}{c_s} \sqrt{\mu - \Delta},
{}~~~~~~~\mu \geq \Delta .
\end{eqnarray}

When Coulomb correlations are taken into account, the kinetic energy of the
holons given by the last term in (5) should be changed by the electrostatic
energy of the Wigner lattice (1).

In the absence of the screening provided by the metallic electrode, the energy
is
given by  Eq.(2), and instead of Eq.(5) we have for the thermodynamic potential
\begin{equation}
\Omega_W (n) = (\Delta - \mu)~n + e^2 n^2 \ln (nL) .
\end{equation}
Comparing (8) with (5) and (7), we find that the square-root singularity of the
compressibility is replaced by a logarithmic singularity:
\begin{equation}
\kappa^{-1}(\mu) \simeq 2 e^2  \ln \frac{(\mu - \Delta) L}{e^2} .
\end{equation}
We have dropped irrelevant numerical factors in the argument of the logarithm.
It is assumed in Eq.(9) that the deviation $\Delta \mu \equiv \mu - \Delta$ is
muchla rger than the Coulomb energy $e^2/L$. One can easily check that, in this
regime, the maximum length of the chain should not exceed
$L_m \sim a \exp (const/n a_B) \gg a$. For a larger length, $L \geq L_m$,
the ordered structure would be destroyed.

The presence of a massive metallic electrode makes the situation more complex.
In the weak screening regime (2), similar to (9), we have a logarithmic
dependence of the compressibility on the density
\begin{equation}
\kappa^{-1}(n) = 2 e^2 \ln (e n D), ~~~n D \geq 1 .
\end{equation}
On further decreasing the density, this dependence is replaced by a power-law,
corresponding to the strong-screening regime: \begin{equation}
\kappa^{-1}(n) = 24 \zeta (3) (e n D)^2 .
\end{equation}
According to Eq.(11), the dependence of the equilibrium
density on chemical potential
is given by
\begin{eqnarray}
n(\mu) &=& 0, ~~~~~~~~~~~~~~~~~~~~~~~~~~~\mu < \Delta^* \\
n(\mu) &=& \frac{1}{2 \lambda}
\left[ \frac{\mu - \Delta^*}{\zeta (3) (e^2 / D)}\right]^{1/3}
{}~~~~~~~\mu \geq \Delta^* ,
\end{eqnarray}
where $\Delta^* = \Delta - e^2 /4 D$. Substituting (13) into (11), one finds
that
the logarithmic singularity (9) of the compressibility, taking place
for unscreened Coulomb potential, is changed to a power-law
$\kappa^{-1} \sim (\mu - \Delta^*)^{2/3}$. However, in the strong screening
regime, at densities $n \leq a_B D^{-2}$, the Wigner lattice is destroyed
locally by quantum fluctuations. Therefore, Eq.(13) is valid at
$\Delta \mu = \mu - \Delta^* \geq (e^2 / D) (a_B /D)^3$.
There is a region near threshold where the universal square-root singularity,
characterizing the commensurate-incommensurate transition, is recovered.
Notice that the massive electrode leads to a shift of the critical value of the
chemical potential, $\mu_c = \Delta - e^2/4D$, caused by the image forces (see
{\em e.g.} [15]).
This shift is small in the case of weak Coulomb coupling
$\alpha_s = e^2/ \hbar c_s \ll 1$. In the strong-coupling regime,
$\alpha_s \geq 1$, one should also account for strong Coulomb interactions
affecting the insulating phase.

For strong Coulomb coupling, $\alpha_s \geq 1$, the characteristic (Bohr)
radius $a_B$ turns out to be shorter than the correlation length $\xi_0$,
indicating the possibility of strong renormalization of the characteristics of
single solitons by Coulomb interaction. From physical considerations one
expects that, due to strong Coulomb interaction, solitons become heavier and
extended. In other words, the effective correlation length $\xi_s$, being the
size of the ``Coulomb" soliton, may substantially esceed $\xi_0$. We shall
study
this situation assuming that screening is extremely small, $\lambda \leq
\xi_0$.

A consistent solution of this problem suggests introducing the long-range
Coulomb interaction at the level of the Hubbard model and calculation of the
resulting spectrum of the charge excitations in such a system. However, for a
half-filled system, repulsive Coulomb effects can only increase the charge gap.
For this reason, reduction of the problem to the sine-Gordon model still
remains reasonable. First we shall study the effect of Coulomb forces on
topological solitons of the quasiclassical sine-Gordon model. Then we we shall
present arguments explaining the applicability of the obtained results to the
holon dynamics in the Hubbard model.

In the presence of long-range Coulomb forces, the Lagrangian of
the sine-Gordon model has the form:
\begin{eqnarray}
{\cal L} = \frac{\hbar}{c_s}
\left[ \frac{1}{2} \dot{\varphi}^2  - \frac{c^2 _s}{2}(\varphi ')^2
+ \frac{\omega^2 _0}{\beta^2} ~(cos \beta \varphi - 1) \right]\nonumber\\
- \frac{e^2 \beta^2}{8 \pi^2} \int_{-\infty}^{\infty}
dy~ \frac{\partial_x \varphi(x,t) \partial_y \varphi(y,t)}
{\sqrt{(x - y)^2 + \lambda^2}} .
\end{eqnarray}
We shall be interested in topological solitons (kinks) of the model (14).
At $\beta^2 \ll 1$ the model is quasiclassical, and the excitations we are
interested in can be found by using trial functions.

A trial function, describing a static topological soliton
$\Delta \varphi \equiv \varphi(x=+\infty) - \varphi(x=-\infty) = 2\pi /\beta$
of the model Eq.(14), will be chosen in the form corresponding to the
sine-Gordon model:
\begin{equation}
\varphi_C (x) = \frac{4}{\beta}
\arctan \left(\exp \left(\frac{x - x_0}{d^*} \right)\right) .
\end{equation}
Here $x_0$ is the center of the soliton, and $d^*$ is a variational parameter
which determines the soliton size. Its value is found by minimizing the soliton
rest energy:
\begin{eqnarray}
E(d^*) &=& \frac{\hbar}{c_s} \int_{-\infty}^{\infty} dx~
\left[ \frac{c^2 _s}{2} (\partial_x \varphi_C)^2
+ \frac{\omega^2 _0}{\beta^2} (1 - \cos \beta \varphi_C) \right]\nonumber\\
&+& \frac{e^2 \beta^2}{8 \pi^2} \int \int dx~dy~
\frac{\partial_x \varphi_C (x) \partial_y \varphi_C (y)}
{\sqrt{(x - y)^2 + \lambda^2}} .
\end{eqnarray}

Substituting (15) into (16) and doing the integrals, we easily find that, for a
weak coupling $\alpha_s \ll 1$, the parameter $d^*$ coincides with the size of
an
unperturbed kink,
$d^* = d_0 = c_s / \omega_0$, $E(d^*) = E_s = 8 \hbar \omega_0 /\beta^2$.
For strong coupling,  $\alpha_s \geq 1$, the unscreened Coulomb interaction
leads to multiplicative renormalizations of the size and energy of the soliton:
\begin{eqnarray}
d^* = d_0 \left(\alpha_s \frac{\beta^2}{2 \pi^2}\right)^{1/2}
\ln ^{1/2} \left[ \frac{d_0}{\lambda} \sqrt{\alpha_s \frac{\beta^2}{2 \pi^2}}
\right] \gg d_0 \\
E_C = \frac{8 \hbar \omega_0}{\beta^2} \left(\alpha_s \frac{\beta^2}{2 \pi^2}
\right)^{1/2}
\ln ^{1/2} \left[ \frac{d_0}{\lambda} \sqrt{\alpha_s \frac{\beta^2}{2 \pi^2}}
\right] \gg E_0 .
\end{eqnarray}

Let us show that expression (18) can be obtained within a consistent scheme,
 using solutions of the sine-Gordon equations. For a linearized problem, the
Coulomb interaction can be taken into account exactly, leading to a
modification
of the spectrum of small perturbations (``optical phonons")
of the field $\varphi$ (see, {\it e.g.} [9])
\begin{equation}
\omega^2 (k) = \omega^2 _0 +
k^2 c^2 _s ~[~ 1 + \alpha_s \frac{\beta^2}{2 \pi^2}K_0 (k \lambda)~ ] .
\end{equation}
Since at $x \geq 1$ $K_0 (x)$ is exponentially small, the Coulomb interaction
mostly affects the long-wavelength dynamics, where the spectrum (19) takes the
form
\begin{equation}
\omega^2 (k) = \omega^2 _0 + k^2 c^2 _s ~\frac{\alpha_s \beta^2}{2 \pi^2}
{}~\ln \frac{1}{k \lambda}, ~~~k \lambda \ll 1 .
\end{equation}

Expression (20) differs from the ``phonon" spectrum of the unpertubed
sine-Gordon equation in that the constant velocity $c_s$ has been replaced by a
momentum dependent effective ``velocity"  $ c_s (k) \sim c_s \sqrt{\ln (1/|k|
\lambda)}$. Analysing the structure of the last term in Eq.(16), one concludes
that, when studying the influence of weakly screened ($\lambda \ll d_0$)
Coulomb  interaction
on the topological soliton of the sine-Gordon model, it is sufficient to change
the exact Lagrangian Eq.(14) by an approximate one in which the effects of
Coulomb interaction are incorporated in a coordinate dependent velocity
\begin{equation}
c^2 _s \rightarrow c^2 (x - x_0) = c^2 _s \frac{\alpha_s \beta^2}{2 \pi^2}
\ln \frac{|x - x_0|}{\lambda} .
\end{equation}
In Eq.(21) it is assumed that $|x - x_0| \gg \lambda$. For this reason, it is
possible
to neglect in all calculations terms containing derivatives of the ``velocity",
$|c'(x)/c(x)| \ll 1$.

In the framework of such an ``adiabatic" perturbation theory, the topological
soliton has the standard form:
\begin{equation}
\varphi_C (x) = \pm \frac{4}{\beta} \arctan [\exp \left(\frac{x - x_0}{d(x)}
\right)],
\end{equation}
where the soliton size, $d_0$, is replaced by a coordinate dependent, smooth
function $d(x) \equiv c(x)/\omega_0$. One can easily check that the energy of
the Coulomb soliton (22) exactly coincides with Eq.(18).

Using the solution (22), the standard scheme of quasiclassical quantization can
be easily developed (see, {\it e.g.} [16]).
It can be shown that the one-loop quantum correction to the classical energy of
the Coulomb soliton has the same form as in the usual sine-Gordon
model, $\Delta E_s = -\hbar \omega_0 /\pi$. Within the
traditional scheme of quasiclassical quantization of solitons, the relative
smallness of quantum corrections is provided by the small value of the coupling
constant $\beta^2 \ll 1$. In our case, solitons can be treated classically due
to large Coulomb energy. Therefore, it seems natural to suggest that formula
(18) remains valid in the extreme quantum  case $\beta^2 \gg 1$ (from the point
of view of the usual sine-Gordon model).

When the charge sector of the half-filled, weakly repulsive ($U/t \ll 1$)
Hubbard model is mapped onto the  sine-Gordon model, the coupling constant
$\beta^2$ turns out to be equal $8\pi$ [11].
The above analysis allows one
to put forward a hypothesis that the long-range Coulomb interaction in the
strong-coupling case $e^2/ \hbar c_s \gg 1$ lead to a multiplicative
renormalization of the Mott-Hubbard gap
\begin{equation}
\Delta \rightarrow \Delta_C = \Delta \sqrt{\frac{e^2}{\hbar c_s}}
{}~\ln ^{1/2} \left( \frac{\xi_0 }{\lambda}
\sqrt{\frac{e^2}{\hbar c_s}}\right) .
\end{equation}
Notice that the appearance of fractional powers of a large logarithm is typical
for various quantum problems that explicitly incorporate long-range Coulomb
forces  [13,9,17].

Now we consider Coulomb solitons in the conduction band at $\mu > \Delta_C$.
At low densities $n \xi_C \ll 1$, where
\begin{equation}
 \xi_C \simeq \xi_0 \alpha^{1/2}_s \ln^{1/2} \left(\frac{\xi_0}{\lambda}
\sqrt{\alpha_s} \right)
\end{equation}
is the characteristic size of the Coulomb soliton, the charges in the
conduction band can be considered as point-like. The unscreened Coulomb
interaction leads to the formation of a Wigner crystal (see Eqs.(2),(9)). On
increasing the density $n \geq\xi^{-1}_C \ll a^{-1}_B$, the solitons start to
overlap strongly, forming a sine-Gordon lattice.

The energy density of such a classical lattice can be readily estimated, using
well-known periodic solutions of the sine-Gordon model
\begin{equation}
\varphi_p (x) = \frac{1}{\beta}
\{ \pi + 2 am~[\frac{x}{d^*} \frac{1}{k(n)} ]  \} .
\end{equation}
Here $am(z)$ is the elliptic amplitude, and $k(n)$ is the elliptic modulus
fixed by the soliton density
\begin{equation}
2 k K (k) = (n d^*)^{-1} ,
\end{equation}
where $K(k)$ is the complete elliptic integral of the {\it I-}order,
and $d^*$ is defined in Eq.(17). The energy density of the soliton lattice
equals (see also [18])
\begin{equation}
E_p (n) = E_C \frac{n}{k} \left\{ E(k) - \case{1}{2} (1 - k^2) K(k) \right\}
\end{equation}
($ E(k)$ is the complete elliptic integral of the {\it II-}order, $E_C$
being the energy of the Coulomb soliton). Expressions (25)-(27) are valid in
our case at densities $n d^* \geq 1$. In the limit of a dense soliton system
$n d^* \gg 1$ $(k \rightarrow 0)$, the soliton lattice smoothly transforms into
a charge-density wave
\begin{equation}
E_p (n) \simeq \frac{\pi^2}{4} (\Delta_s d^*) n^2
\end{equation}
and the compressibility of the system is no longer dependent on the density:
$\kappa^{-1} \sim \Delta_c d^* = const$.

In conclusion we have shown that long-range Coulomb forces
drastically modify the properties of  1D electron systems in
 vicinity of the metal-insulator phase transition. In the metallic
phase a weakly-screened Coulomb interaction leads to the formation of a Wigner
crystal of charged quasiparticles and therefore affects the critical
behaviour of the system at the transition point. The properties
of the insulating phase are also changed if the  short
wavelength screening length is sufficiently small. In this case the
Mott-Hubbard gap is strongly renormalized and the charged
excitations in the Mott phase can be described as quasiclassical
Coulomb solitons.

We gratefully acknowledge discussions with B. L. Altshuler and
K. A. Matveev. This work was supported by the Swedish Royal Academy of
Sciences, the Swedish Natural Science Research Council, and by Grant U2K000
from
the International Science  Foundation. I.K. and A.N.
acknowledge the hospitality of the Department of Applied Physics, CTH/GU.

\newpage

\begin{center}
{\bf REFERENCES}
\end{center}

\noindent [1] P.~W.~Anderson, {\em Science} {\bf 235}, 1196 (1987). \\
\noindent [2] C.~A.~Stafford, and A.~J.~Millis, {\em Phys. Rev.} {\bf B48},
1409 (1993).\\
\noindent [3] H.~J.~Schulz, {\em Int. J. Mod. Phys.} {\bf B5}, 57 (1991). \\
\noindent [4] V.~L.~Pokrovsky, and A.~L.~Talapov, {\em Phys. Rev. Lett.}
{\bf 42}, 65 (1979). \\
\noindent [5] G.~I.~Japaridze, and A.~A.~Nersesyan, {\em JETP Lett.}
{\bf 27},  334 (1978).  \\
\noindent [6] H.~J.~Schulz, {\em Phys. Rev.} {\bf 22}, 5274 (1980). \\
\noindent [7] I.~E.~Dzyaloshinskii, {\em Zh. Eksp. Teor. Fiz.}
{\bf 46}, 1420 (1964).  \\
\noindent [8] E.~Wigner, {\em Phys. Rev.} {\bf 46}, 1002 (1934).  \\
\noindent [9] H.~J.~Schulz, {\em Phys. Rev. Lett.} {\bf 71}, 1864 (1993). \\
\noindent [10] E.~H.~Lieb, and F.~Y.~Wu, {\em Phys. Rev. Lett.} {\bf 20},
1445 (1968). \\
\noindent [11] J.~Solyom, {\em Adv. Phys.} {\bf 28}, 209 (1979). \\
\noindent [12] F.~D.~M.~Haldane, {\em J.Phys. A: Math. Gen.}  {\bf 15}, 507
(1982). \\
\noindent [13] L. I. Glazman, I. M. Ruzin, and B.~I.~Shlovskii,
{\em Phys. Rev.} {\bf B45}, 8454 (1992). \\
\noindent [14] I.~V.~Krive, R.~I.~Shekhter, S.~M.~Girvin, and M.~Jonson,
{\em G{\"o}teborg preprint} APR 93-38 (1993); \\
\noindent [15] M.~Jonson: In {\em ``Quantum Transport in Semiconductors"},
Eds. D.~K.~Ferry, and C.~Jacoboni, Plenum Press, New York (1992). \\
\noindent [16] R. Rajaraman,
{\em Solitons and Instantons: An Introduction to Solitons \\
 and Instantons in Quantum Field Theory},
North Holland, Amsterdam  (1982). \\
\noindent [17] M.~Fabrizio, A.~O.~Gogolin, and S.~Scheidl,
{\em Phys. Rev. Lett.} {\bf 72}, 2235 (1994). \\
\noindent [18] I. V. Krive, and A. S. Rozhavsky,
{\em Int. J. Mod. Phys.} {\bf B6}, 1255 (1992).

\end{document}